\documentclass[epj]{svjour}
\usepackage{graphics}
\newcommand{\be}{\begin{equation}}
\newcommand{\ee}{\end{equation}}

\newcommand{\ba}{\begin{eqnarray}}
\newcommand{\ea}{\end{eqnarray}}

\newcommand{\el}{^}
\begin{document}
\title{Schwarzschild generalized black hole horizon and the embedding space}
\author{J. M. Hoff da Silva\inst{1} \and Rold\~ao da Rocha\inst{2}
}                     
\offprints{}          
\institute{Departamento de F\1sica e Qu\1mica, Universidade
Estadual Paulista, Av. Dr. Ariberto Pereira da Cunha, 333,
Guaratinguet\'a, SP, Brazil. \and Centro de Matem\'atica, Computa\c c\~ao e Cogni\c c\~ao, Universidade Federal do ABC 09210-170, Santo Andr\'e, SP, Brazil.}
\date{Received: date / Revised version: date}
%
\abstract{
By performing a Taylor expansion along the extra dimension of a metric describing a black hole on a brane, we explore the influence of the embedding space on the black hole horizon.  In particular, it is shown that the existence of a Kottler correction of the black hole on the brane, in a viable braneworld scenario, might represent the radius of the black string  collapsing to zero, for some point(s) on the black string axis of symmetry along the extra dimension. Further scrutiny on such black hole corrections by braneworld effects is elicited, the well-known results in the literature are recovered as limiting cases, and we assert and show that when the radius of the black string transversal section is zero, as one moves away from the brane into the bulk, is indeed a  singularity. 
} 
\maketitle

\section{Introduction}
\label{intro}

The intricacy regarding  the study of the gravitational collapse on the brane is consensual. Nonlocal gravitational interaction (between the bulk and the brane) and high-energy corrections as well are just some examples of problems to be dealt with, in the search for analytical solutions \cite{analitico1,analitico2}. A particular exact analysis on the gravitational collapse on the brane in \cite{bb1} can be extended to a general collapse, based on the AdS/CFT correspondence \cite{bb3,bb2}. The numerical approach in the bulk \cite{num1} -- rather difficult to implement -- as well as perturbations  in the background \cite{pert} set by the Randall-Sundrum model \cite{RS2} are among the attempts to study the problem. Nevertheless, there is no solution representing a realistic black hole on the brane, which is  stable and devoid of a  singularity.

An interesting method, based upon a multipole expansion, was developed in \cite{Ita} to probe the shape of a static spherically symmetric braneworld metric along the extra dimension. By the same token, it is possible to use the covariant approach to the braneworld geometry \cite{jap,ali}, exploring the variation of the black hole area horizon along the transverse dimension. More specifically, the covariant formalism allows a well behaved Taylor expansion of the metric about the brane \cite{maar}. It brings prominent information on the relationship between the static gravitational field in the bulk and on the brane as well.   

In this paper we investigate the  Taylor expansion of the brane metric along the extra dimension, based upon the covariant formalism.  Furthermore, it leads to the possible configurations of a (generalized) Schwarzschild black hole area on the brane. We are not devoted, at least in principle, to the particularization to some specific braneworld model. Instead, the approach  aforementioned is implemented without referring to any model \emph{a priori}. Clearly, it is one of the main goals in the covariant approach. Hence, after a general discussion, physical constraints are carried out on the parameters, in order to investigate a more familiar case. 

Within a viable range of parameters, a Kottler-like correction to the brane black hole is obtained. It  might lead to a singularity along the extra dimension: a point in the bulk, representing the black string radius, whose $g_{\theta\theta}$ component equals zero. We prove here that there is a point where the black string radius goes to zero, which is indeed a  singularity: the Kretschmann scalar diverges in such point. This paper is organized as follows: after briefly reviewing the brane field equations in the next Section, in Section 3 the Taylor expansion regarding the brane metric along the extra transverse dimension is introduced. The generalized Schwarzschild black hole horizon is whereafter investigated along the extra  dimension. In Section 4 we analyze the possibility of obtaining a region in the bulk, in a direction orthogonal to the brane, wherein the black string 
presents a radius equal to zero. It is executed by  investigating the black string associated to the mass shifted 
Schwarzschild brane black hole. Two distinct cases, where the bulk is considered as being $dS_5$ or $AdS_5$, are scrutinized. Section 5 is devoted to show the existence of a braneworld model providing a Kottler-like spherically symmetric brane metric. Furthermore, we show that the possibility of the black string 
to present a radius equal to zero, at some point(s) along the extra dimension, may not be excluded, at least  in
principle. At such points, the black string radius equal to zero is shown to be in addition a  singularity. In Section VI  we conclude.

\section{A short review of the basic formalism} 

The fundamental equations concerning the covariant formalism of the gravitational field on the brane are quite well established. Therefore, solely a brief account on it is going to be provided. The expansion procedure, however, is still source of some misunderstanding in the literature. Hence,   to work out the expressions carefully has paramount importance.

By taking the sources of the gravitational field to be  the brane and an unspecified cosmological constant term, the bulk is assumed to satisfy the Einstein field equations
\be \el{(5)}G_{\mu\nu}=-\Lambda_{5}\; \el{(5)}g_{\mu\nu}+\kappa_{5}\el{2}T_{\mu\nu}, \label{1} \ee where $\el{(5)}G_{\mu\nu}$ denotes the Einstein tensor, $T_{\mu\nu}$ is the bulk stress tensor and $\Lambda_{5}$ is the bulk cosmological constant. Moreover, $\kappa_{5}\el{2}$ represents the gravitational coupling in five dimensions and $\el{(5)}g_{\mu\nu}$ is the bulk metric. It is related to the brane metric, $g_{\mu\nu}$, by $\el{(5)}g_{\mu\nu}=g_{\mu\nu}+n_{\mu}n_{\nu}$, where $n_{\mu}$ is an unitary vector normal to the brane. The brane is placed at a fixed point $y=0$, where $y$ hereon denotes the extra dimension. The line element reads 
\be \el{(5)}ds\el{2}=g_{\mu\nu}(x,y)dx\el{\mu}dy\el{\nu}+dy\el{2} \label{2}.\ee The brane and its matter content are present in the definition of $T_{\mu\nu}$ by\footnote{Note that the delta appearing in the definition of $T_{\mu\nu}$ is already conveniently defined in the curved space, since the line element (\ref{2}) has $g_{yy}=1$.} 
\be T_{\mu\nu}=S_{\mu\nu}\delta(y), \ee where $S_{\mu\nu}=-\lambda g_{\mu\nu}+\tau_{\mu\nu}$.  Here $\lambda$ denotes the brane tension while $\tau_{\mu\nu}$ stands for any additional matter on the brane.  
 A prominent quantity that appears in the description of the embedding of manifolds --- the extrinsic curvature --- is given by \be K_{\mu\nu}=g_{\mu}\el{\alpha}g_{\nu}\el{\beta}\nabla_{\alpha}n_{\beta},\label{3} \ee where $\nabla$ denotes the covariant derivative compatible to $\el{(5)}g_{\mu\nu}$. The extrinsic curvature ``measures'' how the brane is embedded into the bulk. Its jump across the brane is provided by the brane stress tensor by the Israel's junction condition \cite{Isr} \be [K_{\mu\nu}]=-\kappa_{5}\el{2}\Big(S_{\mu\nu}-\frac{1}{3}g_{\mu\nu}S\Big),\label{4} \ee where $[\chi]\equiv \lim_{y\rightarrow (0\el{+})}\chi - \lim_{y\rightarrow (0\el{-})}\chi$, for any tensorial quantity $\chi$.   

Following the standard projection procedure \cite{jap} and the $Z_{2}$ symmetry on the left hand side of Eq. (\ref{4}) as well, it is possible to arrive at the induced field equations on the brane \be R_{\mu\nu}-\frac{1}{2}g_{\mu\nu}R =-\Lambda g_{\mu\nu}+\kappa_4\el{2}\tau_{\mu\nu}+\frac{6\kappa_4\el{2}}{\lambda}\pi_{\mu\nu}-E_{\mu\nu},\label{5}\ee where $\kappa_4\el{2}=\frac{\lambda \kappa_{5}\el{4}}{6}$ and \ba\Lambda&=&\frac{1}{2}(\Lambda_{5}+\kappa_4\el{2}\lambda),\label{6} \\ \pi_{\mu\nu}&=&\frac{1}{12}\tau\tau_{\mu\nu}-\frac{1}{4}\tau_{\mu\sigma}\tau\el{\sigma}_{\nu}+\frac{1}{24}(3\tau_{\sigma\delta}\tau\el{\sigma\delta}-\tau\el{2})\label{7} \ea and $E_{\mu\nu}$ is a specific contraction of the bulk Weyl tensor ($\el{(5)}C\el{\alpha}_{\beta\rho\sigma}$) \be E_{\mu\nu}\equiv \,\el{(5)}C\el{\alpha}_{\beta\rho\sigma}n_{\alpha}n\el{\rho}g_{\mu}\el{\beta}g_{\nu}\el{\sigma}.\label{8} \ee

\section{The Taylor expansion}

After having condensed just some results about the projection scheme, we  delve hereon into the main tools necessary to provide the expansion along the extra dimension. The set of Eqs.(\ref{5}) are well known not to perform a closed system, being necessary to study the behavior of $E_{\mu\nu}$. From the geometric point of view the equations ruling the Weyl tensor term may be obtained from Eq.(\ref{1}) and the five-dimensional Bianchi identities $\nabla_{[\mu}\,\el{(5)}R_{\nu\alpha]\beta\sigma}=0$. These equations are used to the Taylor expansion around the brane.  By denoting the Lie derivative along the extra dimension as\footnote{One could denote more precisely  the Lie derivative as $\mathcal{L}_n$, where $n$ is a vector normal to the brane. Nevertheless we opt to a simpler notation $\mathcal{L}$.} $\mathcal{L}$ we have
\ba g_{\mu\nu}(x,y)&=& g_{\mu\nu}(x,0)+[\mathcal{L}g_{\mu\nu}(x,y)]|_{(y=0)}|y|\nonumber\\
&&+[\mathcal{L}[\mathcal{L}(x,y)g_{\mu\nu}(x,y)]]_{(y=0)}\frac{|y|\el{2}}{2!}\nonumber\\&&+\cdots +
[\mathcal{L}\el{k}g_{\mu\nu}(x,y)]_{(y=0)}\frac{|y|\el{k}}{k!}+\cdots .\label{9}\ea Since in our context the brane is indeed described by a manifold, it is quite convenient to emphasize the use of synchronous coordinates --- see Eq. (\ref{2}) --- which makes it possible that the Lie derivative along the transverse dimension may be understood as the usual derivative $\partial/\partial y$.  
Now it is possible to look at each term of Eq. (\ref{9}). The first term after the metric on the brane $g_{\mu\nu}(x,0)$ leads to the definition of the extrinsic curvature \be K_{\mu\nu}=\frac{1}{2}\mathcal{L}g_{\mu\nu}.\label{10}\ee Together with the junction condition (\ref{4}) with $Z_{2}$ symmetry, it completely determines this part of the expansion. The second term is naturally proportional to $\mathcal{L}K_{\mu\nu}$, which is related to the five-dimensional Riemann tensor by \cite{jap,wald} \be \mathcal{L}K_{\mu\nu}=K_{\mu\alpha}K\el{\alpha}_{\,\;\nu}-\el{(5)}\!R_{\mu\alpha\nu\beta}n\el{\alpha}n\el{\beta},\label{11} \ee where the term $\el{(5)}\!R_{\mu\alpha\nu\beta}$ is expressed in terms of the Weyl tensor by  
\ba \el{(5)}\!R_{\mu\alpha\nu\beta}&=&\frac{2}{3}\Big(\,\el{(5)}\!g_{\mu[\nu}\,\el{(5)}\!R_{\beta]\alpha}-\,\el{(5)}\!g_{\alpha[\nu}\,\el{(5)}\!R_{\beta]\mu}\Big)\nonumber\\&&-\frac{1}{6}\,\el{(5)}\!g_{\mu[\nu}\,\el{(5)}\!g_{\beta]\alpha}\,\el{(5)}\!R+\,\el{(5)}\!C_{\mu\alpha\nu\beta}.\label{12} \ea With the aid of Eq. (\ref{1}), where the bulk part does not involve $T_{\mu\nu}$, it is possible to express Eq. (\ref{12}) as 
\be \el{(5)}\!R_{\mu\alpha\nu\beta}=\,\el{(5)}\!C_{\mu\alpha\nu\beta}-\frac{\Lambda_{5}}{9}\,\el{(5)}\!g_{\mu[\nu}\,\el{(5)}\!g_{\beta]\alpha}-\frac{4\Lambda_{5}}{9}\,\el{(5)}\!g_{\alpha[\nu}\,\el{(5)}\!g_{\beta]\mu}.\label{13}\ee Therefore, substituting Eq. (\ref{13}) in the last term of Eq. (\ref{11}) it reads\footnote{Note that $\el{(5)}\!C_{\mu\alpha\nu\beta}n\el{\alpha}n\el{\beta}=\,\el{(5)}\!C\el{\alpha}_{\beta\rho\sigma}n_{\alpha}n\el{\rho}g_{\mu}\el{\beta}g_{\nu}\el{\sigma}$.} \be \mathcal{L}K_{\mu\nu}=K_{\mu\alpha}K_{\mu}\el{\alpha}-\frac{\Lambda_{5}}{6}g_{\mu\nu}-E_{\mu\nu}.\label{14} \ee
 Eqs. (\ref{10}) and (\ref{14}), together with the junction condition supplemented by the $Z_{2}$ symmetry, completely determines the first two $y$-dependent terms of the expansion (\ref{9}):
\ba g_{\mu\nu}(x,y)&=&\left. g_{\mu\nu}(x,0)-\kappa_{5}\el{2}\Bigg[S_{\mu\nu}+\frac{1}{3}(\lambda-S)g_{\mu\nu}(x,0)\Bigg]|y|\right.\nonumber\\&&+\left.\Bigg[-E_{\mu\nu}+\frac{\kappa_{5}\el{4}}{4}\Bigg(S_{\mu\alpha}S\el{\alpha}_{\nu}+\frac{2}{3}(\lambda-S)S_{\mu\nu}\Bigg)\right.\nonumber\\&+&\left.\frac{1}{6}\Bigg(\frac{\kappa_{5}\el{4}}{6}(\lambda-S)\el{2}-\Lambda_{5}\Bigg)g_{\mu\nu}(x,0) \Bigg]|y|\el{2}+\cdots\right.\label{15} \ea 
Particularizing to the brane vacuum case, Eq. (\ref{5}) implies that \be E_{\mu\nu}=-R_{\mu\nu}+\Lambda g_{\mu\nu}.\label{17} \ee 

In order to complete the necessary set up, one notices that the static spherical metric on the brane is \be g_{\mu\nu}(x,0)dx\el{\mu}dx\el{\nu}=-F(r)dt\el{2}+\frac{dr\el{2}}{H(r)}+r\el{2}d\Omega\el{2}.\label{16} \ee The Schwarzschild black hole is well known to be implemented when $F(r)=H(r)= 1-2M/r$ in Eq. (\ref{16}). Now, its generalization is 
accomplished by defining the deviation from the pure Schwarzschild metric as $\psi(r)$, we have $H(r)=1-\frac{2M}{r}+\psi(r)$. Hence, from  the study of the $g_{\mu\nu}(x,y)$ component in Eq. (\ref{15}) it is possible to use \ba \hspace*{-1.5cm}E_{rr} &=& \left(1-\frac{2M}{r}+\psi(r)\right)^{-1}\left(\frac{\psi^{\prime\prime}(r)}{2}+\frac{\psi^\prime(r)}{r}-\Lambda\right)\label{e11}\\ E_{tt} &=& -\left(1-\frac{2M}{r}+\psi(r)\right)\left(\frac{\psi^{\prime\prime}(r)}{2}+\frac{\psi^\prime(r)}{r}-\Lambda\right)\label{e22}\\E_{\theta\theta}&=&\frac{d(r\psi)}{dr}+\Lambda r\el{2}, \qquad E_{\phi\phi} = r^2\sin\theta\; E_{\theta\theta}.\label{e33}\end{eqnarray}\noindent Consequently, Eq. (\ref{15}) has the following non trivial possibilities: 
\ba g_{tt}(x,y)&=&-\left(1-\frac{2M}{r}+\psi\right)\left[1+\frac{2\sqrt{6(2\Lambda-\Lambda_{5})}}{3}|y|\right.\nonumber\\&-&\left.\left(\frac{\psi^{\prime\prime}}{2}+\frac{\psi^\prime}{r}+\frac{5\Lambda_{5}-2\Lambda}{6}\right)|y|\el{2}+\cdots\right] \label{191}\\
 g_{rr}(x,y)&=&\left(1-\frac{2M}{r}+\psi\right)^{-1}\left[1-\frac{2\sqrt{6(2\Lambda-\Lambda_{5})}}{3}|y|\right.\nonumber\\&+&\left.\left(\frac{\psi^{\prime\prime}}{2}+\frac{\psi^\prime}{r}+\frac{25\Lambda_{5}+49\Lambda}{3}\right)|y|\el{2}+\cdots\right] \label{192}\\
g_{\theta\theta}(x,y)&=&r\el{2}\left[1-\frac{2\sqrt{6(2\Lambda-\Lambda_{5})}}{3}|y|\right.\nonumber\\&+&\left.\left(\frac{2\Lambda-5\Lambda_{5}}{6}-\frac{1}{r^2}\frac{d(r\psi)}{dr}\right)|y|\el{2}+\cdots\right] \label{19}\\
g_{\phi\phi}(x,y)&=&(r\el{2}\sin^2\theta)\;g_{\theta\theta}(x,y).\label{193}\ea\noindent
 One can realize forthwith that when $y=0$ --- namely, on the brane --- all the components above are led to their respective values $g_{\mu\nu}(x,0)$ on the brane.  Eqs.(\ref{191}, \ref{192}) and (\ref{19}) are very illustrative, since they explicitly evince: a) the respective departure from 
the metric coefficients on the brane, which are recovered when $y=0$; b) the respective metric corrections, as one move onwards the extra dimension. Such corrections depend merely upon the function $\psi(r)$ (indeed its first  and second order derivatives) and the brane and the bulk cosmological constants.

In the rest of the paper, the generalized Schwarzschild black hole horizon is investigated along the extra transverse dimension, exploring the $g_{\theta\theta}(x,y)$ component in the vacuum brane case. 
Indeed,  the Schwarzschild black hole solution (namely, the black string solution \emph{on the brane}) has the horizon given by $\sqrt{g_{\theta\theta}(x,0)} = r$. Therefore, in the analysis regarding the black string behavior along the extra dimension, we are concerned merely about the horizon behavior, which is provided uniquely by the value for the metric on the brane $g_{\theta\theta}(x,0) = r^2$, defining the radial coordinate.  More specifically, the black string
 horizon ---  or warped horizon \cite{m1} --- is defined for $r = 2M$, which corresponds to the black hole horizon on the brane.
 
It is important to emphasize that, as we shall see, it is possible to find out points along the extra dimension for which $g_{\theta\theta}=0$ and show that in such points the Kretschmann scalar $K =  {}^{(5)}R_{\mu\nu\rho\sigma} {}^{(5)}R^{\mu\nu\rho\sigma}$ diverges, i. e., they are indeed  singularities. Let us remark two points about Eq. (\ref{19}) above. First, we decide to exhibit it in terms of the bulk and the brane cosmological constants. It is convenient in what follows in the next Section. Second and most important, by keeping the sign of the second term as in Eq. (\ref{15}) we are solely considering the case regarding a positive brane tension.  

\section{General cases for the pure black string}

This Section is a prelude to the most interesting analysis of the next Section. As mentioned previously, in this Section we shall consider a Taylor expansion up to the second order in the extra dimension, without assuming any specific model \emph{a priori}. Since the model is not completely specified, it is difficult to pinpoint every physical parameter entering the covariant formalism. This apparent complication, however, is important to study some interesting variations of the current five-dimensional models. In order to assure a physically
 viable analysis, it is implicitly assumed (somewhat tautologically) that the expansion takes place in an infinitesimal region around the brane, such that including the terms up to $y^2$ in Eq.(\ref{9}) suffices.  In the next Section, our study is particularized to some familiar arrange of parameters. Therefore, this question can be further developed. 

The point to be addressed hereon is whether there are points along the extra dimension leading to $g_{\theta\theta}(x,y)=0$. In other words, we aim to investigate whether it is possible to combine the model parameters, as well as the brane black hole correction, in a way that under such assumption, there is a specific  point, let us say $y = y_0$ in the bulk,
along the axis of symmetry of a cylindrical coordinate system, such that  the black string transversal section has radius equal to zero. Therefore, by Eq.(\ref{19}) it follows that the black string squared radius $g_{\theta\theta}(x,y)$ equals zero when
{\ba |y|&=&\left.\frac{r}{\frac{r\el{2}}{6}(2\Lambda-5\Lambda_{5})-\frac{d(r\psi)}{dr}}\Bigg[\pm\sqrt{\frac{d(r\psi)}{dr}+r\el{2}\left(\frac{\Lambda_{5}}{6}+\Lambda\right)}\right.\nonumber\\&&\left.\qquad\qquad\qquad\qquad\qquad\quad+\frac{r}{3}\sqrt{6(2\Lambda-\Lambda_{5})}\Bigg].\right.\label{20} \ea}

The simplest case is the pure black string: the pure Schwarzschild brane black hole $(\psi =0)$ or the mass shifted Schwarzschild $(\psi \sim 1/r)$. It is well known \cite{CHR} that in this case some of its  Kretschmann scalar $K =  {}^{(5)}R_{\mu\nu\rho\sigma} {}^{(5)}R^{\mu\nu\rho\sigma}$ diverges. Indeed,  it is often used to identify singularities, and for a Schwarzschild black hole, $K \propto 1/r^6$, so there is a singularity at $r=0$, but not at the Schwarzschild horizon $r=2M$ \cite{maar}. Besides, the pure black string configuration is unstable \cite{GL}. Hence, this structure is non-physical ab initio. Certainly, for $y=0$ we reproduce the Kretschmann scalars behavior. 
We shall delve into the pure Schwzraschild black string case, in order to fix our procedure. For this particularization,  two distinct cases hold, namely: an arbitrary $dS_{5}$ bulk and an arbitrary $AdS_{5}$ bulk. Let us briefly examine these two cases separately.
 
{For the black string in an arbitrary $dS_{5}$ case, the situation is quite simple. Eq.(\ref{20}) gives
 \be |y|=\frac{6}{2\Lambda-5\Lambda_{5}}\Bigg[\pm\sqrt{\Lambda_{5}/6+\Lambda}+\frac{1}{3}\sqrt{6(2\Lambda-\Lambda_{5})}\Bigg].\label{21} \ee Thus, the cases $\Lambda=0$ and $\Lambda<0$ are immediately ruled out. In addition, the case $\Lambda>0$ is taken into account, since for a real $|y|$ it is necessary $\Lambda_{5}>-\Lambda/6$, which always holds, since  and $2\Lambda>\Lambda_{5}$ must also hold. Furthermore, such conditions are compatible to (\ref{6}). }

{ In the case of a black string in an arbitrary $AdS$ five-dimensional bulk, Eq. (\ref{20}) reads simply \be |y|=\frac{3}{2\Lambda+5|\Lambda_{5}|}\Bigg[\pm\sqrt{-|\Lambda_{5}|/6+\Lambda}+\frac{1}{6}\sqrt{6(2\Lambda+|\Lambda_{5}|)}\Bigg]. \label{22}\ee The square roots above are both real, provided that $|\Lambda_{5}|/3<2|\Lambda|<|\Lambda_{5}|$, which is also appealing from Eq.(\ref{6}). Note that for the positive root, for example, we have a positive $|y|$ if $|\Lambda|<5|\Lambda_{5}|/2$.}

Although a possible combination of the cosmological constants leading to black string with radius zero has been found, as mentioned, this is not the most interesting case. Therefore, let us study a black hole on the brane in the case where the function $\psi(r)$ does not vanish.

\section{The Kottler-like horizon example}

It is useful to investigate the more restrictive case where the effective cosmological constant vanishes \cite{maar}. This is the most appealing case, since it reproduces at first the general assumption $\Lambda \sim 0$. Let us then suppose that the derivative of the correction function $\psi(r)$ does not vanish. In fact, as generalizations of the Schwarzschild solution on the brane with a non-vanishing (bulk) cosmological constant are investigated, we shall look for a Kottler-like line element \cite{KT}. 

Eq.(\ref{6})  forthwith evinces that the brane cosmological constant vanishes when $\Lambda_{5}=-|\Lambda_{5}|$, with $|\Lambda_{5}|=\kappa_{5}\el{4}\lambda\el{2}/6$. Now solely the two cases at Eq. (\ref{20}) are possible.
Taking into account  the positive root it reads 
{ \be |y|=\frac{r}{-\frac{d(r\psi)}{dr}+\frac{5r\el{2}|\Lambda_{5}|}{6}}\Bigg[\sqrt{\frac{d(r\psi)}{dr}-\frac{r\el{2}|\Lambda_{5}|}{6}}+\frac{r}{3}\sqrt{6|\Lambda_{5}|}\Bigg].\label{23}\ee}

 By defining \be \frac{d(r\psi)}{dr}=\beta \frac{r\el{2}|\Lambda_{5}|}{6},\label{24} \ee where $1\leq \beta<2$, it corresponds to a positive and real $|y|$. It might lead in principle to a collapse of the black string  along the extra dimension. However, by substituting the definition (\ref{24}) into Eq. (\ref{23}) it follows that { \be |y|=\frac{1}{(5-\beta)\kappa_{5}\el{2}\lambda}(\sqrt{\beta-1}+2).\label{25} \ee }
 
 At this point it is interesting to conceive an order of magnitude. Since $\kappa_{5}\el{2} \sim 1/M_{5}\el{3}$, where $M_{5}$ is the five-dimensional Planck mass, we have $\kappa_{5}\el{2}\lambda \sim \kappa_{4}\el{2}M_{5}\el{3}$. Just as an example,  the one brane Randall-Sundrum model \cite{RS2} is analyzed, where table-top experiments \cite{tt} probe submillimetric scales of deviations in Newton's law in four dimensions. It provides a lower bound on the five-dimensional Planck mass as $M_{5} > 10\el{5}$ $TeV$. Therefore $\kappa_{5}\el{2}\lambda$ is expected to have a small value. In this vein, in order to have an infinitesimal $|y|$ --- where a second order Taylor expansion is justified --- it would be necessary $\beta$ much greater than $1$, which is prevented by the constraint $1\leq \beta<2$. It is important to  remark on the utilization of such limits. As mentioned, it was obtained by the study of deviations from the Newton's law. Therefore, it concerns weak field experiments which does not does invalidate anything for the case considered. These experiments may be used, indeed, in order to refine the parameters in the model. One shall, however, be aware and not confuse the possible correction provided by $\psi$ with the weak field correction. In fact, keeping in mind the own nature of the problem --- i. e., the black hole horizon along the extra dimension --- $\psi$ is not expected to coincide to any weak field correction.   

The another possibility is given by the negative root in Eq. (\ref{20}). In this case 
{ \be |y|=\frac{r}{-\frac{d(r\psi)}{dr}+\frac{5r\el{2}|\Lambda_{5}|}{6}}\Bigg[-\sqrt{\frac{d(r\psi)}{dr}-\frac{r\el{2}|\Lambda_{5}|}{6}}+\frac{r}{3}\sqrt{6|\Lambda_{5}|}\Bigg],\label{26} \ee}
{
 and using (\ref{24}), and requiring that $-\frac{d(r\psi)}{dr}+\frac{5r\el{2}|\Lambda_{5}|}{6}>0$ together with $-\sqrt{\frac{d(r\psi)}{dr}-\frac{r\el{2}|\Lambda_{5}|}{6}}+\frac{r}{3}\sqrt{6|\Lambda_{5}|}>0$ in order to get a positive (and real) $|y|$, we obtain a case similar to the one in the previous paragraph. In this case, however, the last possibility given by \be -\frac{d(r\psi)}{dr}+\frac{5r\el{2}|\Lambda_{5}|}{6}<0 \label{27}\ee  and \be -\sqrt{\frac{d(r\psi)}{dr}-\frac{r\el{2}|\Lambda_{5}|}{6}}+\frac{r}{3}\sqrt{6|\Lambda_{5}|}<0 ,\label{28}\ee is satisfied by \be \frac{d(r\psi)}{dr}=\alpha \frac{r\el{2}|\Lambda_{5}|}{6},\label{29} \ee provided that $ \alpha > 1.$ Note that substituting Eq. (\ref{29})  in (\ref{26}), and requiring $|y|<1$ we have \be \frac{6(\sqrt{3\alpha-1}-2)}{(3\alpha-5)\kappa_{5}\el{2}\lambda}<1.\label{31} \ee}
 
 Hence, for this case the problem related to the range of the parameter $\alpha$ does not manifest. In fact, for an acceptable result it is necessary that {\be 6(\sqrt{3\alpha-1}-2)<(3\alpha-5)\kappa_{5}\el{2}\lambda,\label{32} \ee} which is satisfied for a huge $\alpha$, of about $\alpha>\frac{12}{\kappa_{5}\el{4}\lambda\el{2}}$. These considerations are going to be more precise in the near future. Before, one observes that Eq. (\ref{29}) provides a Kottler-like correction as \be \psi(r)=\frac{\alpha|\Lambda_{5}|r\el{2}}{6}.\quad \label{33}\ee The reason for the definition (\ref{24}) and (\ref{29}) is obvious, since both lead to a Kottler correction. It is indeed expected, since although the effective cosmological constant vanishes, the bulk cosmological constant acts gravitationally on the brane. Moreover, the opposite sign according to the standard Kottler problem is due to the $AdS$ bulk character.

Eq. (\ref{33}) is shown to supply a type of correction to the Schwarschild brane black hole, which leads (via Eq. (\ref{26})) to the black string radius to be zero, at some point $y_0$ along the extra dimension. It is conceivable to ask about the possibility of such a correction occur. In other words, is  there a braneworld model leading to a Kottler-like spherically symmetric brane line element? The answer is positive, and that is another reason  of Eq. (\ref{29}). This type of configuration may be achieved in any dimension \cite{DD}. It is also possible to construct a Kottler-like spherically symmetric brane line element, by allowing the existence of different bulk cosmological constants in different sides of the brane \cite{CE}. 

At this point it is necessary to analyze the typical constraints on the parameters, checking once again their compatibility to the Taylor expansion. Note first that $\kappa_{5}\el{2}\lambda=\frac{3}{4\pi}\kappa_{4}\el{2}M_{5}\el{3}$ and $\kappa_{4}\sim 10\el{-16}$ $(TeV)\el{-1}$. Recall that $M_{5}$ is constrained by table-top experiments in such way that we may introduce a dimensionless parameter $b>1$ such that $M_{5}=b\,10\el{5}$ $TeV$. In terms of $b$, it is immediate to see that $\kappa_{5}\el{2}\lambda=\frac{3}{4\pi}b\el{3}10\el{-17}$ $TeV$. Moreover, from the constraint imposed on $\alpha$ (see below Eq. (\ref{32})) it follows that $\alpha>\frac{4\el{3}\pi\el{2}}{3b\el{6}}10\el{34}$. Inserting a new scale, say $\tilde{a}=\frac{3b\el{6}}{4\el{3}\pi\el{2}}a>1$, we have simply $\alpha=a\,10\el{34}$ $(TeV)\el{-2}$, where $a$ has dimension of $(TeV)\el{2}$. 

Likewise, it is also important to discuss a little further about the viability of the correction obtained in the light of the constraints above. From Eq. (\ref{33}) it is straightforward to see that the general requirement that the first term must be bigger than the second one implies \be \kappa_{5}\el{2}\lambda r\el{2}|y|>\frac{\kappa_{5}\el{2}\lambda}{8}\left(\alpha+\frac{7}{2}\right)\kappa_{5}\el{2}\lambda r\el{2}|y|\el{2}.\label{n1} \ee In the context of a huge $\alpha$ it reads \be \alpha\kappa_{5}\el{2}\lambda|y|<8.\label{n2} \ee Expressing $\alpha$ as before, the constraint (\ref{n2}) means that $|y|<\frac{8\times 10\el{-18}}{a}$ $TeV$. Hence, the plausible imposition $l_{p}<|y|$, where $l_{p}\sim 10\el{-16}$ $(TeV)\el{-1}$ is the Planck length, gives the following upper bound on the parameter $a$: \be a<8\times 10\el{-2} \hspace{.1cm} (TeV)\el{2}.\label{n3}\ee It is clearly suitable, since all the constraints may be satisfied by an appropriate value for the  parameter $b$. To finalize, the ranges of parameters in the expansion shall not be fixed in principle, and the final word concerning the physical viability of the correction (\ref{33}) shall be established by (table-top) experiments, since one ends with a dependence on the $b$ scale.   

For the sake of completeness, it is necessary to point out that in spite of the pragmatic aspect of an  expansion up to the second order in $y$, the coefficient of the third term brings additional contribution. It should be investigated in order to assure that the expansion performed is not in jeopardy. In fact, the next term in the expansion is proportional to \be \mathcal{L}(K_{\mu\alpha}K\el{\alpha}_{\nu})-\mathcal{L}(E_{\mu\nu})-\frac{1}{6}\Lambda_{5}\mathcal{L}(g_{\mu\nu}).\label{16} \ee The first term may be easily obtained using the Leibniz rule together with the junction condition and the last term was already mentioned (\ref{10}). The term in the middle is a little bit more subtle. Defining the `magnetic' part of the Weyl tensor by $B_{\mu\nu\alpha}\equiv g_{\mu}\el{\rho}g_{\nu}\el{\sigma}\,\el{(5)}\!C_{\rho\sigma\alpha\lambda}n\el{\lambda}$, it is possible to show that \cite{jap} \ba \mathcal{L}(E_{\mu\nu})=D\el{\lambda}B_{\lambda(\mu\nu)}+\frac{\kappa_{5}\el{2}\Lambda_{5}}{6}(K_{\mu\nu}-g_{\mu\nu}K)+K\el{\alpha\beta}R_{\alpha\mu\beta\nu}\nonumber\\\hspace{-1cm}+3K\el{\lambda}_{(\mu}E_{\nu)\lambda}-KE_{\mu\nu}+(K_{\mu\sigma}K_{\nu\lambda}-K_{\mu\nu}K_{\sigma\lambda})K\el{\sigma\lambda},\label{17l}\nonumber \ea where  $D_{\mu}$ stands for the metric compatible covariant derivative. Now, taking every contribution into account, the third term of the expansion (in the brane vacuum) is written as 
\be \frac{1}{81}\Bigg[2\lambda\el{3}\kappa_{5}\el{6}r\el{2}-\kappa_{5}\el{6}\lambda\el{2}\alpha r\el{2}\Bigg]|y|\el{3},\label{n41}\ee \noindent which may be discarded when compared to the second term of the expansion.

As we have shown,  the existence of a point $y_0$ in the extra dimension, in which the black string radius equals zero, may not be excluded in principle. For the cases here pointed --- the pure black string ($\psi(r) = 0$), the shifted mass black string ($\psi(r) \sim 1/r$), and the Kottler case --- we calculate the Kretschmann scalar  $K =  {}^{(5)}R_{\mu\nu\rho\sigma} {}^{(5)}R^{\mu\nu\rho\sigma}$ 
taking into account the metric coefficients in Eqs.(\ref{191}, \ref{192}, \ref{19}, \ref{193}). For such cases  the Kretschmann scalar $K$ diverges at $r\rightarrow 0$, and also $K$ diverges at $y=y_0$, irrespective of the value for $r$, characterizing indeed a  singularity. To summarize, the point $y_0$ along the extra dimension which corresponds to the radius of the black string equal to zero is indeed a  singularity. Furthermore, for the case where the cosmological constant $\Lambda = 0$, our results are in full compliance with \cite{num1}. In fact, in such particular case we have  
  $K = {}^{(5)}R_{\mu\nu\rho\sigma} {}^{(5)}R^{\mu\nu\rho\sigma} \propto {48G^2M^2
\over r^6}\, e^{4|y|/\ell}$, which tends to infinity as $r\to 0$. As usual, the parameter $\ell$ is associated with the bulk curvature radius and corresponds
to the effective size of the extra dimension probed by a five-dimensional graviton \cite{maar}. Our results are more general, as we take into account Eqs.(\ref{e11}, \ref{e22}, \ref{e33}).

\section{Final remarks and outlook}

In this paper we investigated which type of  correction in the horizon of a Schwarzschild black hole in the brane would lead to the collapse of the black string, namely, the existence of a coordinate $y_0$ at the extra dimension wherein the black string radius is exactly zero, at such point.  It was accomplished with the aid of a second order Taylor expansion, endowed with the equations coming from the (covariant) projection scheme. It was shown that in the most viable case, namely, the positive tension 3-brane with vanishing effective cosmological constant embedded into a $AdS_{5}$ bulk, it is possible, in principle, to find out a correction leading to a vanishing black string radius along the extra dimension. This correction is nothing but the Kottler one, and it is known that some approaches may lead to such a solution on the brane, making important the study of this particular correction. Hence, at least classically, this problem persists. For all the cases investigated heretofore -- the pure black string, the shifted mass black string, and the Kottler one as well -- the Kretschmann scalar diverges both at $r\rightarrow 0$, and, further, at $y=y_0$ (corresponding to  the point along the extra dimension wherein  the radius of the black string equal to zero) is indeed a  singularity. We shall finalize, however, stressing that the Kottler correction seems not to be feasible. The reason rests upon the reported quantum instability of the quantum vacuum state with non divergent stress energy \cite{WHIS}. We are currently investigating this possibility and believe that the analysis performed in ref. \cite{WHIS} can be {\it mutatis mutandis} applied to the present case.

\section*{Acknowledgments}
J. M. Hoff da Silva is grateful to CNPq
grant 482043/2011-3, and R. da Rocha is grateful to Conselho Nacional de Desenvolvimento Cient\'{\i}fico e Tecnol\'ogico (CNPq)  grants 476580/2010-2 and
304862/2009-6 for financial support.

\end{document}